\documentclass[11pt]{article}
\usepackage{moriond,epsfig}

\def\be{\begin{equation}}
\def\ee{\end{equation}}
\def\bea{\begin{eqnarray}}
\def\eea{\end{eqnarray}}

\begin{document}
\vspace*{4cm}
\title{JET PHYSICS AT THE TEVATRON}

\author{ Sally Seidel \\
{\sl for the CDF and D0 Collaborations}}

\address{Department of Physics and Astronomy, University of New Mexico,\\
800 Yale Blvd. NE\\
Albuquerque, NM 87131, USA}

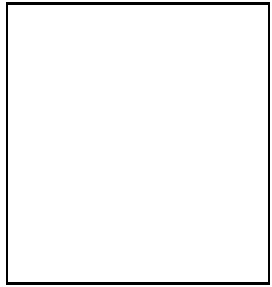
\maketitle\abstracts{Recent analyses by the CDF and D0 Collaborations of
jet data produced in $p\overline p$
collisions at the Fermilab Tevatron Collider
 are presented.
 These include new studies of the inclusive jet production cross section, a
measurement
 of the strong coupling constant, the first measurement of subjet multiplicity
 of quark and gluon jets, examination of ratios of multijet cross 
 sections and their implications for choice of renormalization scale,
 and a study of charged
 jet evolution and energy flow in the underlying event.  The results are
 compared to theoretical predictions.}

\section{Jets at CDF and D0}

Jets have been studied by the CDF~\cite{cdf} and D0~\cite{d0}
Collaborations as a means of
searching for new particles and interactions, testing a variety of 
perturbative QCD predictions, and providing input for the global parton
distribution function fits.   Unless otherwise indicated below, the jets
were reconstructed~\cite{cdfjets,d0jets}
using a cone algorithm with cone radius $R=0.7$ from data taken at the
Fermilab Tevatron collider in 1994-95 with 
$\sqrt{s}=1.8$ TeV.

\section{Measurement of the Inclusive Jet Production Cross Section (CDF)}

A new measurement~\cite{cdfincl} has been
made of the inclusive jet cross section
using 87 pb$^{-1}$ of data.  This cross section, which probes a distance
scale of order
$10^{-17}$ cm, stimulated interest in obtaining improved precision of
parton distribution functions (PDFs) when initial measurements showed an excess
of data over theoretical predictions at high transverse energy ($E_T$).
The analysis uncouples the systematic shift in the cross section
associated with energy ``smearing'' (the combined effects of energy
mismeasurement and resolution limitation in the detector) from the
statistical uncertainty on the
data by comparing observed data to theoretical predictions that have been
``smeared.''  A $\chi^2$ fitting 
technique was developed to quantify the degree to which the next-to-leading
order (NLO) calculation~\cite{EKS} agrees with data for each PDF examined.  An
additional technique was developed to determine the probability
distributions associated with the $\chi^2$ variable.  The CTEQ4HJ PDF was
found to have the highest confidence level (10\%), followed by MRST with
7\%.  CDF Run 1b data were found to be consistent with CDF Run 1a data,
with D0 results, and
with NLO QCD, given the flexibility allowed by current knowledge of PDFs.

\section{The Pseudorapidity and Transverse Energy Dependence of the
Inclusive Jet Production Cross Section (D0)}

A measurement~\cite{d0eta} of the pseudorapidity ($\eta$) and $E_T$
dependence of the
inclusive jet cross section using 95 pb$^{-1}$ of data has extended the
kinematic range of this quantity considerably beyond previous measurements
by examining the differential cross section for $|\eta|$ up to 3.  A
correlation matrix technique was used
to compare the data to the NLO calculation~\cite{JETRAD} for several PDFs.
The data indicate a preference for the CTEQ4HJ, MRSTg$\uparrow$, and CTEQ4M
PDFs.

\section{Measurement of the Strong Coupling Constant from Inclusive Jet
Production (CDF)}

A measurement~\cite{alphas} of the strong coupling constant,
$\alpha_s(M_Z)$, was extracted
from the inclusive jet cross section by comparing the measured
cross section in each of 33 $E_T$ bins to a NLO prediction~\cite{JETRAD}
using available matrix elements~\cite{matrix} and evolving each of the
resultant
$\alpha_s$ values to $M_Z$.  The data show good agreement with QCD for 
$40 < E_T < 250$ GeV.  The value of $\alpha_s$ averaged over that range is
$0.1178 \pm 0.0001 ({\rm stat.}) ^{+0.0081}_{-0.0095} ({\rm exp.\ syst.})$.
The extracted values $\alpha_s(M_Z)$ are $E_T$-independent and consistent
with those input to the PDFs.  The principal theoretical uncertainties
are due to choice
of renormalization scale $\mu_R$ $(^{+6\%}_{-4\%})$ and input PDF (5\%).

\section{The Inclusive Jet Cross Section using the $k_\perp$ Algorithm (D0)}

The inclusive jet cross section was measured,~\cite{d0incl} for
87 pb$^{-1}$ of data
and $|\eta|<0.5$,
using jets reconstructed with the $k_\perp$ algorithm.~\cite{d0kt}  While
the NLO predictions~\cite{JETRAD} with the $k_\perp$ and cone algorithms
are consistent within 1\%, the cross section measured with the $k_\perp$
algorithm is 37\% higher than with the cone.  The theoretical prediction
is lower than the data by about 50\% at the lowest transverse momentum
$p_T$ ($\sim 60$ GeV/c) 
and by 10-20\% for $p_T > 200$ GeV/c.  Inclusion of hadronization effects
in the NLO prediction improves the agreement somewhat.  With hadronization
effects considered, the CTEQ4HJ and MRST PDFs show agreement with $\chi^2$
probability of 44\% and 46\%, respectively.

\section{Ratios of Multijet Cross Sections (D0)}

The ratio of the inclusive three-jet to inclusive two-jet production cross
sections as a function of total transverse energy was measured.~\cite{ratio}
The measurements
were used to infer the preferred renormalization scale for modelling
soft jet emission.  For CTEQ4M, no prediction accurately describes the
ratio through the kinematic region; nonetheless a single scale
$\mu_R=0.3\sum{E^{jet}_T}$ adequately describes the rate of additional jet
emission when all uncertainties are considered.  The introduction of additional
scales does not significantly improve agreement with the data.

\section{Subjet Multiplicity of Quark and Gluon Jets Reconstructed with the
$k_\perp$ Algorithm (D0)}

The $k_\perp$ algorithm was used to examine the properties of
jets.~\cite{d0kt}  For
a jet separation parameter $D=1.0$, the jet $p_T$ was found to be higher 
by 5(8) GeV than
the $E_T$ of matching jets reconstructed with cones of radius $R=0.7$
for $p_T\approx 90 (240)$ GeV.  To resolve the internal jet structure
known as subjets, the 
$k_\perp$ algorithm was also applied to $D=0.5$ jets.  Subjet multiplicity
$M$ is expected to depend upon color factor, $\sqrt{s}$, and the 
dimensionless $y_{\rm cut}$ parameter that controls the resolution. 
Because PDF data show that the fraction of gluon jets decreases with 
$x \propto p_T/\sqrt{s}$,
gluon-enriched and quark-enriched datasets may be separated statistically
by selecting jets with the same $p_T$ in two-jet events of $\sqrt{s}=630$~GeV
and $\sqrt{s}=1800$~GeV.  HERWIG was used to predict the gluon fractions $f$,
which in the range $55<p_T<100$ GeV/c were found to be $f_{1800}=0.59$ and
$f_{630}=0.33$.  For gluon multiplicity $M_g$ and quark multiplicity $M_q$
independent of $\sqrt{s}$, the total multiplicity $M$ may be written 
in terms of the multiplicities $M_{1800}$ and $M_{630}$ measured at
$\sqrt{s}=1800$ GeV
and 630 GeV, respectively, as
$M=fM_g+(1-f)M_q$, leading to $M_g={{(1-f_{630})M_{1800}-(1-f_{1800})M_{630}}
\over {f_{1800}-f_{630}}}$ and 
$M_q={{f_{1800}M_{630}-f_{630}M_{1800}}
\over {f_{1800}-f_{630}}}$.  Correction for shower detection effects in the
calorimeter yields subjet multiplicities of $2.21\pm 0.03$ for gluon jets
and $1.69\pm 0.04$ for quark jets.  After unsmearing, the ratio of average
subjet multiplicities in gluon and quark jets is found to be
${{\langle M_g \rangle - 1} \over {\langle M_q \rangle - 1}}=
1.84\pm 0.15(\rm stat.)^{+0.22}_{-0.18}(\rm syst.)$.

\section{Charged Jet Evolution and the Underlying Event (CDF)}

Collider data were compared to predictions by HERWIG, ISAJET, and PYTHIA
in a two-part analysis.~\cite{underlying}
In the first part, observables associated with the
leading charged jet, which provides
information about the hard scatter, including final state radiation, were
studied.  In the second part,
the underlying event, which includes beam-beam remnants, initial state 
radiation, and multiple parton scattering, was studied with global observables.
Minimum bias data and charged jets with $|\eta|<1$ and $0.5<p_T<50$ GeV/c
were examined with a cone algorithm modified for application to low $p_T$
charged particles.  The QCD hard scattering models were found to describe
the leading (highest
$\sum p_T$) charged jet well with regard to charged particle multiplicity,
size, radial and momentum distributions of charged particles, and
$p_T$ around the jet direction.  Charged particle clusters are evident in the
minimum bias data above $p_T \approx 2$ GeV/c.  The global observables,
average charged particle multiplicity $\langle N_{\rm chg}\rangle $ and average
total transverse momentum $\langle \sum p_T \rangle$, are correlated
with the angle relative to the axis of the leading jet, such that
the region transverse to the leading jet (normal to the plane of the
$2\rightarrow 2$ hard scatter) is most sensitive to beam-beam fragments and
initial state radiation.  The $\langle N_{\rm chg}\rangle $ and
$\langle \sum p_T \rangle$ grow rapidly with $p_T^{\rm leading}$ below
5 GeV/c, then plateau.  The plateau height in the transverse
direction is half the height in the direction of the leading jet.
PYTHIA 6.115 models $\langle N_{\rm chg}\rangle $ best in the transverse
region but overestimates it in the direction of the leading jet.  ISAJET
models the amount of jet activity well but does not reflect the
$p_T$ dependence accurately.  Although HERWIG does not model sufficient
$\sum p_T$ in the events, HERWIG and PYTHIA generally model the hard scatter
(especially the initial state radiation) component of the underlying event
best.


\section*{References}
\begin{thebibliography}{99}

\bibitem{cdf} F. Abe, {\sl et al.}, Nucl. Instr. and Meth. A 271, 387 (1988);
F. Abe, {\sl et al.}, Nucl. Instr. and Meth. A 268, 75 (1988).

\bibitem{d0} S. Abachi, {\sl et al.}, Nucl. Instr. and Meth. A 338, 185 (1994).

\bibitem{cdfjets} F. Abe, {\sl et al.}, Phys. Rev. D 45, 1448 (1992);
T. Affolder, {\sl et al.}, Phys. Rev. D 64, 032001 (2001).

\bibitem{d0jets} B. Abbott, {\sl et al.}, Phys. Rev. D 64, 32003 (2001).

\bibitem{cdfincl} T. Affolder, {\sl et al.}, Phys. Rev. D 64, 32001 (2001). 

\bibitem{EKS} S. Ellis, {\sl et al.}, Phys. Rev. Lett. 62, 2188 (1989).

\bibitem{d0eta} B. Abbott, {\sl et al.}, Phys. Rev. Lett. 86, 1707 (2001).

\bibitem{JETRAD} W.T. Giele, {\sl et al.}, Phys. Rev. Lett. 73, 2019 (1994). 

\bibitem{alphas} T. Affolder, {\sl et al.}, Phys. Rev. Lett. 88, 042001 (2002).

\bibitem{matrix} S.D. Ellis, {\sl et al.}, Phys. Rev. Lett. 69, 3615 (1992).

\bibitem{d0incl} V.M. Abazov, {\sl et al.}, Phys. Lett. B 525, 211 (2002).

\bibitem{d0kt}  V.M. Abazov, {\sl et al.}, Phys. Rev. D 65, 052008 (2002).

\bibitem{ratio} B. Abbott, {\sl et al.}, Phys. Rev. Lett. 86, 1955 (2001).

\bibitem{underlying} T. Affolder, {\sl et al.}, Phys. Rev. D 65, 092002 (2002).

\end{thebibliography}
\end{document}